# An $O(n^{2.75})$ algorithm for online topological ordering[*]

Deepak Ajwani[†]   Tobias Friedrich[†]   Ulrich Meyer[†]

**Abstract**

We present a simple algorithm which maintains the topological order of a directed acyclic graph with $n$ nodes under an online edge insertion sequence in $O(n^{2.75})$ time, independent of the number of edges $m$ inserted. For dense DAGs, this is an improvement over the previous best result of $O(\min\{m^{\frac{3}{2}} \log n, m^{\frac{3}{2}} + n^2 \log n\})$ by Katriel and Bodlaender. We also provide an empirical comparison of our algorithm with other algorithms for online topological sorting. Our implementation outperforms them on certain hard instances while it is still competitive on random edge insertion sequences leading to complete DAGs.

## 1 Introduction

A topological order $T$ of a given directed acyclic graph (DAG) $G = (V, E)$ (with $n := |V|$ and $m := |E|$) is a linear ordering of its nodes such that for all directed paths from $x \in V$ to $y \in V$ ($x \neq y$), it holds that $T(x) < T(y)$. There exist well known algorithms for computing the topological ordering of a DAG in $O(m + n)$ in an offline setting (see e.g. [2]).

[*]Research partially supported by DFG grant ME 2088/1-3.
[†]Max-Planck-Institut für Informatik, Stuhlsatzenhausweg 85, 66123 Saarbrücken, Germany

In the online variant of this problem, the edges of the DAG are not known in advance but are given one at a time. Each time an edge is added to the DAG, we are required to update the bijective mapping $T$.

The online topological ordering has been studied in the following contexts

- As an online cycle detection routine in pointer analysis [13].
- Incremental evaluation of computational circuits [1].
- Compilation [8, 10] where dependencies between modules are maintained to reduce the amount of recompilation performed when an update occurs.

The naïve way of maintaining an online topological order, i.e., to compute it each time from scratch with the offline algorithm, takes $O(m^2 + mn)$ time. Marchetti-Spaccamela et al. [9] (MNR) gave an algorithm that can insert $m$ edges in $O(mn)$ time. Alpern et al. proposed a different algorithm [1] (AHRSZ) which runs in $O(\|\delta\| \log \|\delta\|)$ time per edge insertion with $\|\delta\|$ measuring the number of edges of the minimal node subgraph that needs to be updated. Note that not all edges of this subgraph need to be visited and hence even $O(\|\delta\|)$ time per insertion is not optimal. Katriel and Bodlaender (KB) [7] analyzed a variant of the AHRSZ algorithm and obtained an upper bound of $O(\min\{m^{\frac{3}{2}} \log n, m^{\frac{3}{2}} + n^2 \log n\})$ for a general DAG. In addition, they show that their algorithm runs in time $O(m \cdot k \cdot \log^2 n)$ for a DAG for which the underlying undirected graph has a treewidth $k$. Also, they give an $O(n \log n)$ algorithm for DAGs whose underlying undirected graph is a tree. The algorithm by Pearce and Kelly [12] (PK) empirically outperforms the other algorithms for sparse random DAGs, although its worst-case runtime is inferior to KB.

We propose a simple algorithm that works in $O(n^{2.75} \sqrt{\log n})$ time and $O(n^2)$ space, thereby improving upon the results of Katriel and Bodlaender for dense DAGs. With some simple modifications in our data structure, we can get $O(n^{2.75})$ time with $O(n^{2.25})$ space or $O(n^{2.75})$ *expected* time with $O(n^2)$ space. We also demonstrate empirically that this algorithm clearly outperforms MNR, AHRSZ, and PK on a certain class of hard sequences of edge insertions, while being competitive on random edge sequences leading to complete DAGs.

Our algorithm is dynamic, as it also supports deletion. However, our analysis holds only for a sequence of insertions. Our algorithm can also be used for online cycle detection in graphs, as well. Moreover, it permits an arbitrary starting point, which makes a hybrid approach possible, i.e., using the PK or KB algorithm for sparse graphs and ours for dense graphs.

The rest of this paper is organized as follows. In Section 2, we describe the algorithm and the data structures involved. In Section 3, we give the correctness argument for our algorithm, followed by an analysis of its runtime in Sections 4 and 5. The details of our implementation and a empirical comparison with other algorithms follow in Section 6.

## 2 Algorithm

We keep the current topological order as a bijective function, $T : V \to [1..n]$. If we start with an empty graph, we can initialize $T$ with an arbitrary permutation, otherwise $T$ is the topological order of the starting graph, computed offline. In this and the subsequent sections, we will use the following notations: $d(u, v)$ denotes $|T(u) - T(v)|$, $u < v$ is a short form of $T(u) < T(v)$, $u \to v$ denotes an edge from $u$ to $v$, and $u \rightsquigarrow v$ expresses that $v$ is reachable from $u$. Note that $u \rightsquigarrow u$, but *not* $u \to u$.

Figure 1 gives the pseudo code of our algorithm. Throughout the process of inserting new edges, we maintain some data structures which are dependent on the current topological order. Inserting a new edge $(u, v)$ is done by calling INSERT$(u, v)$. If $v > u$, we do not change anything in the current topological order and simply insert the edge into the graph data structure. Otherwise, we call REORDER to update the topological order as well as the data structures dependent on it. As we will prove in Theorem 4, detecting $v = u$ indicates a cycle. If $v < u$, we first collect sorted sets $A$ and $B$ as defined in the code. If both $A$ and $B$ are empty, we swap the topological order of the two nodes and update the data structures. The query and the update operations are described in more detail along with our data structures in Section 2.1. Otherwise, we recursively call REORDER until everything inside is topologically ordered. To make these recursive calls efficient, we first merge the sorted sets $\{v\} \cup A$ and $B \cup \{u\}$ and using this merged list, compute the set $\{u' : (u' \in B \cup \{u\}) \wedge (u' > v')\}$ for each node $v' \in \{v\} \cup A$.

INSERT($u, v$)

 ▷ Insert edge $(u, v)$ and calculate new topological order
1 **if** $v \leq u$ **then** REORDER($u,v$)
2 insert edge $(u, v)$ in graph

REORDER($u, v$)

 ▷ Reorder nodes between $u$ and $v$ such that $v \leq u$
1 **if** $u = v$ **then** report detected cycle and quit
2 $A := \{w : v \to w \text{ and } w \leq u\}$
3 $B := \{w : w \to u \text{ and } v \leq w\}$
4 **if** $A = \emptyset$ and $B = \emptyset$
   **then** ▷ Correct the topological order
5    swap $u$ and $v$
6    update the data structure
   **else** ▷ Reorder node pairs between $u$ and $v$
7    **for** $v' \in \{v\} \cup A$ in decreasing topological order
8     **for** $u' \in B \cup \{u\} \wedge u' \geq v'$ in increasing topological order
9      REORDER($u',v'$)

Figure 1: Our algorithm

## 2.1 Data structure

We store the current topological order, as a set of two arrays, storing the bijective mapping $T$ and its inverse. This ensures that finding $T(i)$ and $T^{-1}(u)$ are constant time operations.

The graph itself is stored as an array of vertices. For each vertex we maintain two adjacency lists, which keep the incoming and outgoing edges separately. Each adjacency list is stored as an array of buckets of vertices. Each bucket contains at most $t$ nodes for a fixed $t$. Depending on the concrete implementation of the buckets, the parameter $t$ is later chosen to be approximately $n^{0.75}$ so as to balance the number of inserts and deletes from the buckets and the extra edges touched by the algorithm. The $i$-th bucket ($i \geq 0$) of a node $u$ contains all adjacent nodes $v$ with $i \cdot t < d(u,v) \leq (i+1) \cdot t$. The nodes of a bucket are stored with node index (and not topological order) as their key. The bucket can be kept as a balanced binary tree or as an array of $n$-bits or as a hash-table of a universal hashing function. The bucket data structure should provide efficient support for the following three operations:

1. Insert: Inserting an element in a given bucket.

2. Delete: Given an element and a bucket, find out if that element exists in that bucket. If yes, delete the element from there and return 1. Else, return 0.

3. Collect-all: Copying all the elements from the bucket to some vector.

Depending on how we choose to implement the buckets, we get different runtimes. This will be discussed in Section 5. We will now discuss how we do the insertion of an edge, computation of $A$ and $B$, and updating the data-structure under swapping of nodes in terms of the above three basic operations.

Inserting an edge $(u,v)$ means, inserting node $v$ to the forward adjacency list of $u$ and $u$ to the backward adjacency list of $v$. This requires $O(1)$ bucket inserts.

For given $u$ and $v$, the set $A := \{w : v \to w \text{ and } w < u\}$ sorted according to the current topological order can be computed from the adjacency list of $v$ by sorting all nodes of the first $\lceil d(u,v)/t \rceil$ outgoing buckets and choosing all $w$ with $w < u$. This can be done by $O\big(d(u,v)/t\big)$ collect-all operations on buckets collecting a total of $O(|A|+t)$ elements. These elements are integers in the range $\{1 \mathinner{\ldotp\ldotp} n\}$ and can be sorted in $O(|A|+t+\sqrt{n})$ time using a two-pass radix sort algorithm. The set $B$ is computed likewise from the incoming edges.

When we swap two nodes $u$ and $v$, we need to update the adjacency lists of $u$ and $v$ as well as that of all nodes $w$ that are adjacent to $u$ and/or $v$. First, we show how to update the adjacency lists of $u$ and $v$. If $d(u,v) > t$, we have to build their adjacency lists from scratch. Otherwise, the new bucket boundaries will differ from the old boundaries by $d(u,v)$ and at most $d(u,v)$ nodes will need to be transferred between any pair of consecutive buckets. The total number of transfers are therefore bounded by $d(u,v)\lceil n/t \rceil$. Determining whether a node should be transferred can be done in $O(1)$ using the inverse mapping $T^{-1}$ and as noted above, a transfer can be done in $O(1)$ bucket inserts and deletes. Hence, updating the adjacency lists of $u$ and $v$ needs $\min\{n, d(u,v)\lceil n/t \rceil\}$ bucket inserts and deletes.

Let $w$ be a node which is adjacent to $u$ or $v$. Its adjacency list needs to be updated only if $u$ and $v$ are in different buckets. This corresponds to $w$ being in different buckets of the adjacency lists of $u$ and $v$. Therefore, the number of nodes to be transferred between different buckets for maintaining the adjacency lists of all $w$'s is the same as the number of nodes that need to be transferred for maintaining the adjacency lists of $u$ and $v$, i.e., $\min\{n, d(u,v)\lceil n/t \rceil\}$.

Updating the mappings $T$ and $T^{-1}$ after such a swap is trivial and can be done in constant time. Thus, we conclude that swapping nodes $u$ and $v$ can be done by $O(d(u,v)\lceil n/t \rceil)$ bucket inserts and deletes.

# 3 Correctness

**Theorem 1.** *The above algorithm returns a valid topological order after each edge insertion.*

*Proof.* For a graph with no edges, any ordering is a correct topological order, and therefore, the theorem is trivially correct. Assuming that we have a valid topological order of a graph $G$, we show that when inserting a new edge $(u, v)$ using INSERT$(u, v)$, our algorithm maintains the correct topological order of $G' := G \cup \{(u, v)\}$. If $u < v$, this is trivial.

We need to prove that $x < y$ for all nodes $x, y$ of $G'$ with $x \rightsquigarrow y$. If there was a path $x \rightsquigarrow y$ in $G$, Lemma 2 gives $x < y$. Otherwise (if there is no $x \rightsquigarrow y$ in G), the path $x \rightsquigarrow y$ must have been introduced to $G'$ by the new edge $(u, v)$. Hence $x < y$ in $G'$ by Lemma 3 since there is $x \rightsquigarrow u \rightarrow v \rightsquigarrow y$ in $G'$. □

**Lemma 2.** *Given a DAG $G$ and a valid topological order. If $u \rightsquigarrow v$ and $u < v$, then all subsequent calls to* REORDER *will maintain $u < v$.*

*Proof.* Let us assume the contrary. Consider the first call of REORDER which leads to $u > v$. Either this call led to swapping $u$ and $w$ with $v \leq w$ or it caused swapping $w$ and $v$ with $w \leq u$. Note that in our algorithm, a call of REORDER$(u, v)$ leads to a swapping only if $A = \emptyset$ and $B = \emptyset$. Assuming that it was the first case (swapping $u$ and $w$) caused by the call to REORDER$(u, w)$, $A = \emptyset$. However, $x \in A$ for an $x$ with $u \rightarrow x \rightsquigarrow v$, leading to a contradiction. The other case is proved similarly.

□

**Lemma 3.** *Given a DAG $G$ with $v \rightsquigarrow y$ and $x \rightsquigarrow u$, a call of* REORDER$(u, v)$ *will ensure that $x < y$.*

*Proof.* The proof follows by induction on the recursion depth of REORDER$(u, v)$. For leaf nodes of the recursion tree, $A = B = \emptyset$. If $x < y$ before this call happens, Lemma 2 ensures that $x < y$ will continue. Otherwise, $y := v$ and $x := u$. The swapping of $u$ and $v$ in line 5 gives $x < y$.

We assume this lemma to be true for calls of REORDER up to a certain tree level. If $A \neq \emptyset$, then there is a $\tilde{v}$ such that $v \to \tilde{v} \rightsquigarrow y$, otherwise $\tilde{v} := v = y$. If $B \neq \emptyset$, then there is a $\tilde{u}$ such that $x \rightsquigarrow \tilde{u} \to u$, otherwise $\tilde{u} := u = x$. Hence $\tilde{v} \rightsquigarrow y < x \rightsquigarrow \tilde{u}$. The **for**-loops of lines 7 and 8 will call REORDER$(\tilde{u}, \tilde{v})$. By the inductive hypothesis, this will ensure $x < y$. According to Lemma 2, further calls to REORDER will maintain $x < y$. $\square$

**Theorem 4.** *The algorithm detects a cycle if and only if there is a cycle in the given edge sequence.*

*Proof.* "$\Rightarrow$": First, we show that within a call to INSERT$(u, v)$, there are paths $v \rightsquigarrow v'$ and $u' \rightsquigarrow u$ for each recursive call to REORDER$(u', v')$. This is trivial for the first call to REORDER and follows immediately by the definition of $A$ and $B$ for all subsequent recursive calls to REORDER. This implies that if the algorithm indicates a cycle in line 1 of REORDER, there is indeed a cycle $u \to v \rightsquigarrow v' = u' \rightsquigarrow u$. In fact, the cycle itself can be computed using the recursion stack of the current call to REORDER.

"$\Leftarrow$": Consider the edge $(u, v)$ of the cycle $v \rightsquigarrow u \to v$ inserted last. Since $v \rightsquigarrow u$ before the insertion of this edge, the topological order computed will have $v < u$ (Theorem 1) and therefore, REORDER$(u, v)$ would be called. In fact, all edges in the path $v \rightsquigarrow u$ will obey the current topological ordering and by Lemma 2, it will remain so for all subsequent calls of REORDER. We prove by induction on the number of nodes in the path $v \rightsquigarrow u$ (including $u$ and $v$) that whenever $v \rightsquigarrow u$ and REORDER$(u, v)$ is called, it detects the cycle. A call of REORDER$(u', v')$ with $u' = v'$ or REORDER$(u', v')$ with $v' \to u'$ clearly reports a cycle. Consider a path $v \to x \rightsquigarrow y \to u$ of length $k > 2$ and the call of REORDER$(u, v)$. As noted before, $v < x \leq y < u$ before the call to REORDER$(u, v)$. Hence $x \in A$ and $y \in B$ and a call to REORDER$(y, x)$ will be made in the for loop of lines 7 and 8. As $y \rightsquigarrow x$ has $k - 2$ nodes in the path, the call to REORDER$(y, x)$ (by our inductive hypothesis) will detect the cycle. $\square$

## 4 Runtime

**Theorem 5.** *Online topological ordering can be computed using $O(n^{3.5}/t)$ bucket inserts and deletes, $O(n^3/t)$ bucket collect-all operations collecting $O(n^2 t)$ elements, and $O(n^{2.5} + n^2 t)$ operations.*

*Proof.* Lemma 7 shows that REORDER is called $O(n^2)$ times. Lemma 9 shows that the calculation of the sets $A$ and $B$ over all calls of REORDER can be done by $O(n^3/t)$ bucket collect-all operations touching $O(n^2 t)$ edges, and $O(n^{2.5} + n^2 t)$ operations. In Lemma 12, we prove that all the updates can be done by $O(n^{3.5}/t)$ bucket inserts and deletes.

As for lines 7 and 8, we first merge the two sorted sets $A$ and $B$, which takes $O(|A| + |B|)$ operations. For a particular node $v' \in \{v\} \cup A$, we can compute the set $V' = \{u' : (u' \in B \cup \{u\}) \wedge (u' > v')\}$ (as required by line 8) using this merged set in complexity $O(1 + |V'|)$, which is also the number of calls of REORDER emanating for this particular node. Summing over the entire *for* loop of line 7, the total complexity of lines 7 and 8 is $O(|A| + |B| + \#(\text{calls of REORDER emanating from here}))$. Since by Lemma 8, the summation of $|A| + |B|$ over all calls of REORDER is $O(n^2)$ and by Lemma 7, the total number of calls to REORDER is also $O(n^2)$, we get a total of $O(n^2)$ operations for lines 7 and 8. Putting everything together, the theorem follows. □

**Lemma 6.** *REORDER is local, i.e., a call to REORDER$(u,v)$ does not affect the topological ordering of nodes $w$ such that either $w < v$ or $w > u$ just before the call was made.*

*Proof.* This theorem can be proved by induction on the level of recursion tree of the call to REORDER$(u,v)$. For the leaf node of the recursion tree, $|A| = |B| = 0$ and the topological order of $u$ and $v$ is swapped, not affecting the topological ordering of any other node.

We assume this lemma to be true up to a certain tree level. To see that it is valid even for a level higher, note that the arrays $A$ and $B$ contain elements $w$ such that $v < w < u$. Since each call of REORDER in the **for**-loop of line 7 and 8 is from an element of $A$ to an element of $B$ and all of these calls are themselves local by our induction hypothesis, this call of REORDER is also local. □

**Lemma 7.** REORDER *is called* $O(n^2)$ *times.*

*Proof.* Let $u$ and $v$ be arbitrary nodes. Let us consider the first time, REORDER$(u,v)$ is called. If $A = B = \emptyset$, $u$ and $v$ will be swapped. Otherwise, REORDER$(u',v')$ is called recursively for all $v' \in \{v\} \cup A$ and $u' \in B \cup \{u\}$ with $u' > v'$. The order in which we make these recursive calls and the fact that REORDER is local (Lemma 6) ensures that REORDER$(u,v)$ is not called except as the last of these recursive calls. In this second call to REORDER$(u,v)$, $A = B = \emptyset$. To see this consider all $v' \in A$ and $u' \in B$ ($A$ and $B$ from the first call of REORDER$(u,v)$). REORDER$(u,v')$ and REORDER$(u',v)$ must have been called within the **for**-loop of the first execution of REORDER$(u,v)$ before this second call was made. Therefore it follows from Lemma 3 and Lemma 2 that before the second call, $u < v'$ and $u' < v$ for all $v' \in A$ and $u' \in B$. Hence $u$ and $v$ will be swapped at the latest in the second call of REORDER$(u,v)$. Since REORDER$(u,v)$ is only called if $v < u$, REORDER$(u,v)$ will not be called again. Hence, REORDER$(u,v)$ is called at most two times for each node pair $(u,v)$. □

**Lemma 8.** *The summation of* $|A|+|B|$ *over all calls of* REORDER *is* $O(n^2)$.

*Proof.* Consider arbitrary nodes $u$ and $v'$. We prove that for all $v \in V$, $v' \in A$ happens only once over all calls of REORDER$(u,v)$. This proves that $\sum |A| \leq n$, for all such calls of REORDER$(u,v)$. Therefore summing up for all $u \in V$, $\sum |A| \leq n^2$ over all calls of REORDER.

In order to see that for all $v \in V$, $v' \in A$ happens only once over all calls of REORDER$(u,v)$, observe that $v' \in A$ implies that $v' < u$ before REORDER$(u,v)$ was called. In particular, $v' < u$ before the call of REORDER$(u,v')$ in the **for**-loop of REORDER$(u,v)$ (follows from the order of recursive calls) and by Lemma 3, $u < v'$ after this call. Therefore, $v' \notin A$ for a call of REORDER$(u,w)$ for any node $w$ after this call. The same is true for all calls of REORDER$(u,w)$ before this call as otherwise $u < v'$ even before the beginning of the current call of REORDER$(u,v)$ and $v' \notin A$ for the current call. Also, $v' \notin A$ for any of the recursive calls of this call to REORDER$(u,v')$. This follows from the order in which we make the recursive calls and the fact that REORDER is local (Lemma 6).

Analogously, it can be proved that for arbitrary nodes $v$ and $v'$ and for all $u \in V$, $v' \in B$ happens only once over all calls of REORDER$(u,v)$. The proof for $\sum |B| \le n^2$ follows similarly and it completes the proof for this lemma. □

**Lemma 9.** *Calculating the sorted sets $A$ and $B$ over all calls of* REORDER *can be done by $O(n^3/t)$ bucket collect-all operations touching a total of $O(n^2 t)$ elements and $O(n^{2.5} + n^2 t)$ operations for sorting these elements.*

*Proof.* Consider the calculation of set $A$ in a call of REORDER$(u, v)$. As discussed before in Section 2.1, we look at the out adjacency list of $u$, stored in the form of buckets. In particular, we will need $O(d(u,v)/t)$ bucket collect-all operations touching $O(|A| + t)$ elements to calculate $A$. The additional worst-case factor of $t$ stems from the last bucket visited. Summing up over all calls of REORDER, we get $O\bigl(\sum d(u,v)/t\bigr)$ collect-alls touching $\sum(|A| + |B| + t)$ elements. Since $d(u,v) \le n$ for every call of REORDER$(u,v)$ and there are $O(n^2)$ calls of REORDER (Lemma 7), there are $O(n^3/t)$ bucket collect-all operations. Also, since $\sum(|A| + |B|) = O(n^2)$ by Lemma 8, the total number of elements touched is $O(n^2 + \sum t) = O(n^2 t)$. Since the keys are in the range $\{1 \mathinner{.\,.} n\}$, we can use a two-pass radix sort to sort the elements collected from the buckets. The total sorting time over all calls of REORDER is $\sum(2(|A| + t) + \sqrt{n}) + \sum(2(|B| + t) + \sqrt{n}) = O(n^{2.5} + n^2 t)$. □

**Lemma 10.** *Each node-pair is swapped at most once.*

*Proof.* REORDER$(u,v)$ is called only when $v < u$. Once a swapping happens, $u < v$. By Lemma 2, it will remain so for all calls of REORDER thereafter. Therefore, REORDER$(u,v)$ is never called again and $u$ and $v$ will not be swapped again. □

**Lemma 11.** *$\sum d(u,v) = O(n^{5/2})$ where the summation is taken over all calls of* REORDER$(u,v)$ *in which $u$ and $v$ are swapped.*

*Proof.* Let $T^*$ denote the final topological ordering and

$$X(T^*(u), T^*(v)) := \begin{cases} d(u,v) & \text{if and when REORDER}(u,v) \text{ leads to a swapping} \\ 0 & \text{otherwise} \end{cases}$$

Since by Lemma 10 any node-pair is swapped at most once, the variable $X(i,j)$ is clearly defined. Next, we model a few linear constraints on $X(i,j)$, formulate it as the linear program and use this LP to prove that $\max\{\sum_{i,j} X(i,j)\} = O(n^{5/2})$. By definition of $d(u,v)$ and $X(i,j)$,

$$0 \leq X(i,j) \leq n \quad \text{for all } i,j \in [1\mathinner{\ldotp\ldotp} n].$$

For $j \leq i$, the corresponding edges $(T^{*\ -1}(i), T^{*\ -1}(j))$ go backwards and thus are never inserted at all. Consequently,

$$X(i,j) = 0 \quad \text{for all } j \leq i.$$

Now consider an arbitrary node $u$, which is finally at position $i$, i.e., $T^*(u) = i$. Over the insertion of all edges, this node has been moved left and right via swapping with several other nodes. Strictly speaking, it has been swapped right with nodes at final positions $j > i$ and has been swapped left with nodes at final positions $j < i$. Hence, the overall movement to the right is $\sum_{j>i} X(i,j)$ and to left is $\sum_{j<i} X(j,i)$. Since the net movement (difference between the final and the initial position) must be less than $n$,

$$\sum_{j>i} X(i,j) - \sum_{j<i} X(j,i) \leq n \quad \text{for all } 1 \leq i \leq n.$$

Putting all the constraints together, we aim to solve the following linear program.

$$\max \sum_{\substack{1 \leq i \leq n \\ 1 \leq j \leq n}} X(i,j) \text{ such that}$$

(i) $X(i,j) = 0$ for all $1 \leq i \leq n$ and $1 \leq j \leq i$

(ii) $0 \leq X(i,j) \leq n$ for all $1 \leq i \leq n$ and $i < j \leq n$

(iii) $\sum_{j>i} X(i,j) - \sum_{j<i} X(j,i) \leq n-1$ for all $1 \leq i \leq n$

Note that these are necessary constraints, but not sufficient. But this is enough for our purpose as an upper bound to the solution of this LP will give an upper bound for the $\sum X(i,j)$ in our algorithm. In order to prove the upper bound on the solution to this LP, we consider the dual problem

$$\min \left[ n \sum_{\substack{0 \leq i < n \\ i < j < n}} Y_{i \cdot n+j} + n \sum_{0 \leq i < n} Y_{n^2+i} \right] \text{ such that}$$

(i) $Y_{i \cdot n+j} \geq 1$ for all $0 \leq i < n$ and for all $j \leq i$

(ii) $Y_{i \cdot n+j} + Y_{n^2+i} - Y_{n^2+j} \geq 1$ for all $0 \leq i < n$ and for all $j > i$

(iii) $Y_i \geq 0$ for all $0 \leq i < n^2 + n$

and the following feasible solution for the dual:

$$\begin{aligned}
Y_{i \cdot n+j} &= 1 && \text{for all } 0 \leq i < n \text{ and for all } 0 \leq j \leq i \\
Y_{i \cdot n+j} &= 1 && \text{for all } 0 \leq i < n \text{ and for all } i < j \leq i+1+2\sqrt{n} \\
Y_{i \cdot n+j} &= 0 && \text{for all } 0 \leq i < n \text{ and for all } j > i+1+2\sqrt{n} \\
Y_{n^2+i} &= \sqrt{n-i} && \text{for all } 0 \leq i < n.
\end{aligned}$$

This solution has a value of $n^2 + 2n^{\frac{5}{2}} + n\sum_{i=1}^{n}\sqrt{i} = O(n^{\frac{5}{2}})$, which by the primal-dual theorem is a bound on the solution of the original LP.

In fact, it can be shown that there is a solution to primal LP whose value is $O(n^{\frac{5}{2}})$, namely

$$X(i,j) = 0 \text{ for all } 0 \leq i < n \text{ and for all } 0 \leq j \leq i$$

$$X(i,j) = n \text{ for all } 0 \leq i < n \text{ and for all } i < j \leq i + \lceil \tfrac{\sqrt{1+8i}-1}{2} \rceil$$

$$X(i,j) = 0 \text{ for all } 0 \leq i < n \text{ and for all } j > i + \lceil \tfrac{\sqrt{1+8i}-1}{2} \rceil.$$

□

**Lemma 12.** *Updating the data structure over all calls of* REORDER *requires* $O(n^{3.5}/t)$ *bucket inserts and deletes.*

*Proof.* Our data structure requires $O(d(u,v)\,n/t)$ bucket inserts and deletes to swap two nodes $u$ and $v$. Each node pair is swapped at most once (cf. Lemma 10). Hence, summing up over all calls of REORDER$(u,v)$ where $u$ and $v$ are swapped, we need $O(\sum d(u,v)\,n/t) = O(n^{3.5}/t)$ bucket inserts and deletes using Lemma 11. □

## 5 Bucket data structure

We get different runtimes and space requirements of our algorithm depending on the data structures of the buckets used:

(a) Balanced binary trees: Balanced binary trees give us $O(1+\log \tau)$ time insert and delete and $O(1+\tau)$ time collect-all operation, where $\tau$ is the number of elements in the bucket. Therefore, by Theorem 5, the total time required will be $O(n^2 t + n^{3.5} \log n/t)$. Substituting $t = n^{0.75}\sqrt{\log n}$, we get a total time of $O(n^{2.75}\sqrt{\log n})$. The total space requirement will be $O(n^2)$ as a balanced binary tree needs $O(t)$ nodes for storing at most $t$ elements.

(b) $n$-bit array: A bucket that stores at most $t$ elements can be kept as an $n$-bit array, where each bit is 0 or 1 depending on whether or not the element is present in the bucket. Also, we can keep a list of all elements in the bucket. To insert, we just flip the appropriate bit and insert at the end of the list. To delete, we just flip the appropriate bit. To collect all, we go through the list and for each element in the list, we check if the corresponding bit is 1 or 0. If it is 0, we also remove it from the list. This gives us constant-time insert and delete and the time for collect-all operation will be the total output size plus the total number of delete. Each delete is counted once in collect-all as we remove the corresponding element from the list after the first collect-all. By Theorem 5, the total time required will be $O(n^2 t + n^{3.5}/t)$, giving us $O(n^{2.75})$ for $t = n^{0.75}$. The total space requirement will be $O(n)$ for each bucket, leading to a total of $O(n^{2.25})$ for $O(n^2/t)$ buckets.

(c) Uniform Hashing [11]: A data structure based on uniform hashing coupled with a list of elements in the bucket operated in the same way as the $n$-bit array will give an expected constant-time insert and delete and the same bound for collect-all as for the $n$-bit array. This gives an expected total time of $O(n^2 t + n^{3.5}/t)$. With $t = n^{0.75}$ this yields an expected time of $O(n^{2.75})$. Since the hashing based data structure as described in [11] takes only linear space, the total space requirement is $O(n^2)$.

# 6 Empirical Comparison

## 6.1 Configuration

We conducted our experiments on a 2.4 GHz Opteron machine with 8GB of main memory running Debian GNU/Linux. For PK, MNR, and AHRSZ we used the C++/Boost based implementation of David J. Pearce [12]. For our algorithm (AFM), we implemented variant (b) of Section 5 using C++/STL. All codes were compiled using gcc 3.3 in 32-bit mode and optimization level "-O3". The timings were measured using the `gettimeofday` function of `<sys/time.h>` and all the results are averaged over 10 runs each.

## 6.2 DAG classes considered

We first consider random edge insertion sequences leading to a complete DAG. For $m < \binom{n}{2}$, this will result in a random DAG, similar to the $G(n,m)$ random graph model of Erdős [3, 4]. On a random edge sequence, all the algorithms are quite fast and none of them encounters its worst-case behavior. Therefore, we consider a particular sequence of edges which we believe is a hard instance of the problem. This edge sequence is similar to the worst-case sequence given by Katriel et al. for their algorithm. On this sequence, PK, MNR and AHRSZ (the variant choosing the smallest permitted priority) face their worst-case of $\Omega(n^3)$ operations, while our algorithm takes $\Omega(n^{2.5})$ time complexity. This sequence of edges is depicted in Fig. 2.

For an example with $n$ nodes, we divide the set of nodes into four blocks of different sizes: block 1 consist of nodes $[0 \ldots n/3)$, block 2 of nodes $[n/3 \ldots n/2)$, block 3 of nodes $[n/2 \ldots 2n/3)$, and block 4 of nodes $[2n/3 \ldots n)$. First, we insert $n-4$ edges such that within each block, the vertices form a directed path from left to right. Then we insert the following edges,

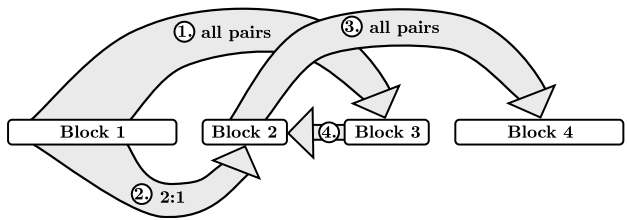

Figure 2: Our hard-case graph

(a) $\overrightarrow{\forall}\, j \in [0..n/3)\ \overleftarrow{\forall}\, k \in [0..n/6) :$ add edge$(j, k + n/2)$,

(b) $\overrightarrow{\forall}\, j \in [0..n/6) :$ add edge$(2j, j + n/3)$ and edge$(2j + 1, j + n/3)$,

(c) $\overrightarrow{\forall}\, j \in [0..n/6)\ \overleftarrow{\forall}\, k \in [0..n/3) :$ add edge$(j + n/3, k + 2n/3)$,

(d) $\overrightarrow{\forall}\, j \in [0..n/6)\ \overleftarrow{\forall}\, k \in [0..n/6) :$ add edge$(j + n/2, k + n/3)$,

where $\overrightarrow{\forall}$ denotes going from left to right in the **for**-loop and $\overleftarrow{\forall}$ the other way around. Similar sequences, which force AHRSZ to encounter its asymptotic worst-case complexity, can be chosen for all variants of AHRSZ.

## 6.3 Results

Fig. 3 shows the runtimes of the four algorithms in consideration for random edge sequences leading to complete DAGs with varying number of vertices $n$ (and with $m = \binom{n}{2}$). We see that AFM is a constant factor of 2-4 away from AHRSZ, MNR and PK.

Fig. 4 shows the average runtimes for random graphs with $n = 1000$ and a varying number of edges. AFM looses a lot in the first $O(n \log n)$ edges because in this phase, updating the data-structures after every swapping proves very costly. But after that, the curves between AFM and PK/MNR are almost parallel, while the slope for AHRSZ is around 2 times that of AFM. For practical purposes, we believe therefore that a hybrid approach would perform best. That is, one inserts the first $O(n \log n)$ edges with either PK or KB and then inserts the remaining edges with our algorithm.

Fig. 5 shows the runtimes of the four algorithms in consideration on the class of hard edge sequences described before. The difference in asymptotic behaviour as discussed before is clear from the graph. For $n = 8000$, AFM is 2 times faster than MNR, 3.6 times faster than PK, and 30 times faster than AHRSZ.

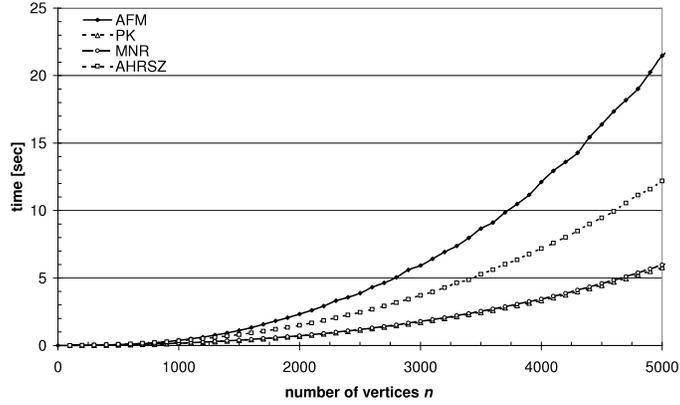

Figure 3: Experimental data on full random graphs with varying $n$

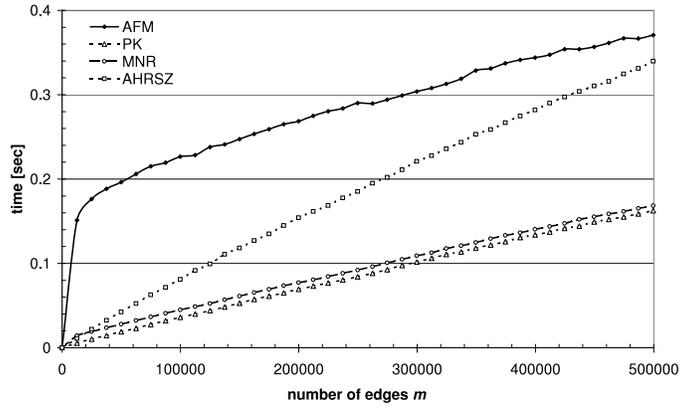

Figure 4: Experimental data on random graphs with $n = 1000$ and varying $m$

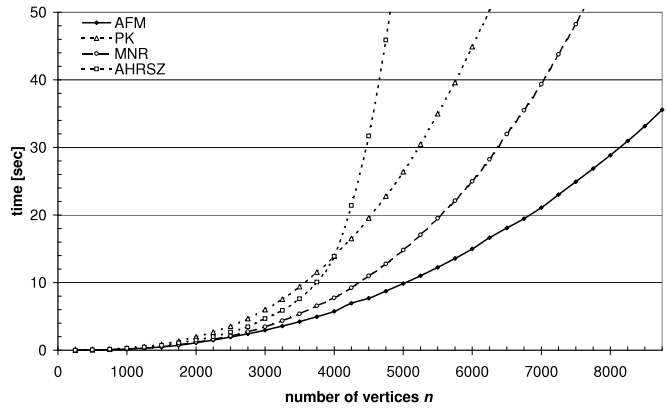

Figure 5: Experimental data on a class of hard instances with varying $n$

# 7 Discussion

We have presented the first $o(n^3)$ algorithm for online topological ordering. We also implemented this new algorithm and compared it with previous approaches, showing that for certain hard examples, it outperforms PK, MNR, and AHRSZ, while it is still competitive on random edge sequences leading to complete DAGs. The only non-trivial lower bound for this problem is by Ramalingam and Reps [14], who show that an adversary can force any algorithm maintaining explicit labels to need $\Omega(n \log n)$ time complexity for inserting $n-1$ edges. There is still a large gap between this, the trivial lower bound of $\Omega(m)$, and the upper bound of $O(\min\{m^{1.5} + n^2 \log n, m^{1.5} \log n, n^{2.75}\})$. Bridging this gap remains an open problem.

# Acknowledgements

The authors are grateful to David J. Pearce for providing us his code. Also, thanks are due to the anonymous referees of SWAT'06 for valuable comments and to Seth Pettie and Saurabh Ray for helpful discussions.